\documentclass[twocolumn,showpacs,prb]{revtex4}
\usepackage{graphicx} %
\usepackage{dcolumn}
\usepackage{amsmath}
\usepackage{latexsym}
\usepackage{longtable}
\begin{document}
\title{ Electrical and magnetic transport study on strain driven ferromagnetic insulating thin film of low doped $\mathrm{La_{1-x}Ca_{x}MnO_3}$  }
\author{Rajib Nath\footnote[1]{email: rajibn@bose.res.in}, Sudeshna Samanta and A. K. Raychaudhuri\footnote[2]{email: arup@bose.res.in}}
\affiliation{Department of Materials Science, S.N.Bose National Centre for Basic Sciences, Block JD, Sector III, Salt Lake, Kolkata 700 098, West Bengal, India.}

\begin{abstract}
\noindent
In this paper we have created a strain driven single crystal like ferromagnetic insulating (FMI) state in a PLD grown thin film of low doped $\mathrm{La_{1-x}Ca_{x}MnO_3}$ (${X\approx 0.15}$)on NGO(100) substrate
 and make a thorough study of strain effects on the electric and magnetic transport of this film. We have studied and compared the FMI state , ferromagnetic transition temperature ($T_{C}$), ferromagnetic insulating temperature ($T_{FMI}$) and the resistivity ($\rho$) of the film in details with bulk single crystals of X=0.18 to 0.22 doping region. We have found that $T_{FMI}$  and the localisation length of the carriers are increased and there is also a decrease in the Coulomb gap. The magneto transport behavior of the film also differs from the bulk single crystals and the magnetoresistance of the sample is nearly ${20\sim75\%}$ with the application of the applied field(${0-10}$ T) and it falls up to ${5\sim 40 \%}$ below a certain temperature and the ${T_{C}}$ of the film increases to higher temperature with the increasing field. The film also shows anisotropic magnetoresistance as large as 20$ \% $ depending on applied magnetic field direction.
\end{abstract}
\pacs{75.47.Lx, 72.70.+m, 73.50.Td, 71.30.+h, 75.30.+h, 75.70.-i}
\maketitle
\section{\bf INTRODUCTION}
Substrate induced strain plays an important role  in the electric and manetic transport behavior of magnetoresistive manganites due to the strong inter-correlation between the spin, charge and lattice degrees of freedom \cite{ref1,ref2}. In the critcal doping region (${X\approx 0.2-0.225}$), CMR manganites ($\mathrm{La_{1-x}Ca_{x}MnO_3}$) shows a ferromagnetic insulating state (FMI) between the paramagnetic insulating (PI) and feromagnetic metallic state(FMM) and generally  magnetoresistance (MR) collapses in this  state\cite{ref3,ref4}. The strain between the film and the substrate also controls the equilibrium between  FMM and FMI staes in  manganites \cite{ref2,ref5}. It has been found that the properties such as the ferromagnetic transition temperature ($ T_{C} $), resistivity(${\rho}$) and MR are sensitive to the epitaxial strain as it can effect the Jahn-Teller electron-phonon coupling \cite{ref6,ref7,ref8}. The effects of epitaxial strain are different from the changes induced by external parameters like hydrostatic or chemical pressure as in-plane strain generally leads to an out-of-plane strain with different sign. The lattice-substrate mismatch  induces phase separation and inhomogeneities in films and thus causes electronic behavior which was not found in bulk materials of the same composition\cite{ref9}. Anisotropic magnetoresistance (AMR) is another important effect in manganites due to its possible application in magnetic memory devices. AMR is mostly observed in the ferromagnetic(FM) state of manganite thin films and strongly depends on the lattice strain, film thickness and the local inhomogeneity of the thin film. This effect has not been explored so much in the $ \mathrm {Ca} $ doped FMI manganites i.e. in the critically doped region\cite{ref10,ref11,ref12,ref13}. So it is very important to study the effect of biaxial strain, micro structure on the electric and magnetic transport behavior of the thin film of manganites in the critically doping region as the nature of the strain (compressive or tensile) can tune the system into more metallic or insulating state\cite{ref8}. The growth of good quality single crystalline thin film retaining FMI state is very important as the lattice strain can make the system FMM instead of being FMI or even more insulating without the FMI state.
In this paper, in order to systematically investigate the electric and magnetic transport of low hole doped  manganite thin films, we have made a single crystal like FMI thin film on $\mathrm{NaGdO_{3}}$ (NGO) substrate from the target of $\mathrm{La_{0.85}Ca_{0.15}MnO_3}$ by pulsed laser deposition (PLD) method . The target composition is selected near to the critical composition $ X=0.225 $ and as we stated earlier that the substrate induced strain can leads the film to the highly insulating to metallic state. FMI and FMM regions  give us more freedom to investigate the strain effect  on transport properties, transition temperatures and phase transitions. In this paper, we have shown in detail experimental investigations of electrical and magnetic transport of this thin film by varying the magnetic field from ${0}$ to ${10}$ Tesla in the temperature (T) range ${50}$ to ${300}$ K. From the measurement we have found that the thin film behaves like a  bulk single crystal sample of manganite in the doping range ${X\approx 0.2-0.22}$ with substantial MR in this state. Lattice strain significantly changes the $ T_{C} $ of the film. There is a substantial AMR in the sample which increases as the applied  field increases.
\section{\bf RESULT and DISCUSSIONS}
 We have used pulsed laser deposition (PLD) system using a eximer laser(KrF, $\lambda$ = 293 nm) for growing the thin film on $\mathrm{NaGdO_{3}}$ (100) single crystal substrate (length 3 mm and width 2 mm) from a 25 mm diameter $\mathrm{La_{0.85}Ca_{0.15}MnO_{3}} $ target pellet. The Laser fluence used for the deposition is $1.6 J/cm^{2}$ and the repeatation rate of the laser pulse was 3 Hz. The substrate was at $ 800^{0}C $ and the surrounding $ O_{2} $ pressure was 0.3 mbar.  
 All the electrical and magnetic transport measurements were done on the thin film of length 3 mm and width 2 mm . For resistivity and magnetoresistance measurements of the sample , we have made four thermally evaporated chrome-gold pad on the film and the four probe electrical contacts are made by attaching Cu wires on the chrome-gold pads using silver paint. A current source (Keithley 220) and an electrometer ( Keithley 6514) were used for the $ \rho $ vs T measurement. A cryogen free magnet was used to apply the magnetic field upto 10 T. For the characterisation of the sample, we have done XRD by Xpert PRO X-ray diffractometer. To avoid the joule heating of the sample ,$1\mu$A  current is sourced for the electrical transport measurement with the current density $ {1.6 A/ cm^{2}} $ .
 \subsection {XRD of the thin film}
   We performed high resolution XRD at room temperature to confirm the phase formation of the film as shown in FIG. 1. The major peak at $32.9 ^{0}$ matches with the NGO substrate and the small peak at $32.3 ^{0}$ is due to the film which is $ 0.3 ^{0} $ less than the standard peak of $\mathrm{La_{0.85}Ca_{0.15}MnO_3}$ and $ 0.6 ^{0} $ less from the standard NGO peak. As the film is deposited on NGO (100), the in plane lattice tends to adopt the same structure as NGO, and the out-of-plane parameter changes correspondingly to maintain unit cell volume. In this process the film will suffer a compressive strain in the XY plane as in plane lattice parameters \textquotedblleft a\textquotedblright and \textquotedblleft b\textquotedblright of NGO are less than the bulk  $\mathrm{La_{0.85}Ca_{0.15}MnO_3}$  while tensile perpendicular to the film since out of plane lattice parameter \textquotedblleft c\textquotedblright is greater than $\mathrm{La_{0.85}Ca_{0.15}MnO_3}$ (NGO: a=5.431 $ \AA{} $, b=5.499 $ \AA{} $, c=7.710 $ \AA{} $ and $\mathrm{La_{0.85}Ca_{0.15}MnO_3}$ :a=5.506 $ \AA{} $, b=7.7465 $ \AA{} $, c=5.4769 $ \AA{} $ ). So the lattice constant \textquotedblleft d\textquotedblright will increase for the film and XRD peak appears at lower angle than the substrate and the bulk $\mathrm{La_{0.85}Ca_{0.15}MnO_3}$. We used the equation $ \epsilon_{ZZ} = \frac{Sin\theta_{substrate}}{Sin\theta_{Film}}-1 $  to calculate strain in Z direction  and fitted the sample and substrate peaks with Gaussian curve. We estimated  $1.8\%$ tensile strain along the perpendicular direction and the lattice constant of the film differs from the bulk $\mathrm{La_{0.85}Ca_{0.15}MnO_3}$ and NGO substrate . The composition of the film from lattice constant value is expected to in the critically doping region of $ X = 0.2-0.225 $. Broadening of the film peak can be attributed to the film thickness which comes out around 30 nm from the data which is in good agreement with our calibrated oxide deposition growth rate in PLD. 
   \begin{figure}[h]
   \begin{center}
   \includegraphics[width=9 cm,height=7 cm]{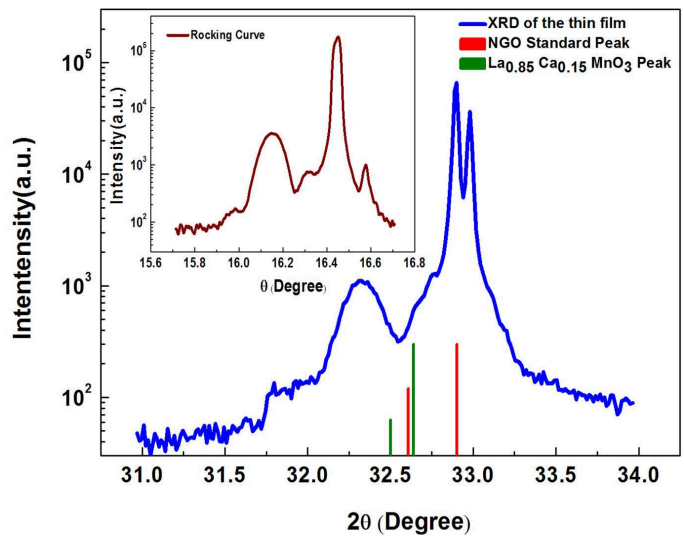}
   \end{center}
   \caption{(Color online) XRD and Rocking Curve (in the inset) of the thin film on NGO substrate}
   \label{Fig1}
   \end{figure}     

 \subsection { Electrical transport of the film}
In order to investigate the electronic transport properties of the film, temperature dependence of resistivity ($ \rho $) was measured on the film from 50 to 300 K as shown in FIG. 2. In the $\rho-T $ curve, there are insulator to metal transition ($T_{C}$) at 181 K and again at 137 K . Ferromagnetic metal to ferromagnetic insulator transition ($T_{FMI}$) occurs and it is the key features of the manganites sample in the low doping region. 
 \begin{figure}[h]
   \begin{center}
   \includegraphics[width=9 cm,height=7cm]{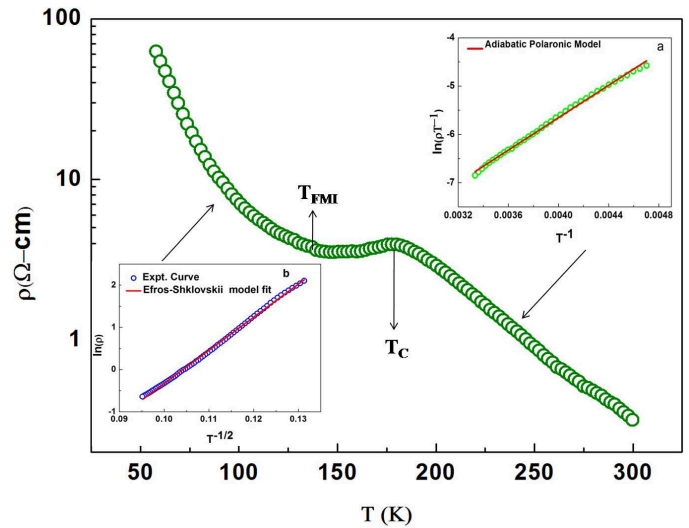}
   \end{center}
   \caption{(Color online)Temperature dependent resistivity of the film. In the inset FIG. a is the Adiabatic polaronic model fit at higher temperature and FIG. b is the Efros-Shlovshkii model fit in the lower temperature range.}
   \label{Fig2}
   \end{figure}      
 The high temperature electrical transport of manganites can be well described by the adiabatic polaronic model($\rho=\rho_{0}T\exp \frac{E_{a}}{\kappa_{B}T}$)\cite{ref14,ref15} and we have fitted the $\rho-T$ data with this model in the temperature range 208 to 300 K (inset a) and we get the activation energy ($ E_{a} $) = 135 meV with $\rho_{0}$ as $4.32\times 10^{-6}$. It is well established that electrical transport in the FMI state follows Efros-Shlovskii variable range hopping (ESVRH) model \cite{ref16,ref17}. We fitted $\rho-T $ data from 58 to 130 K  with the ESVRH model i.e. $\rho = \rho_{0}\exp (\frac{T_{0}}{T})^{1/2}$. From the fitting, we get $ T_{0} $ = 6000 K and the Coulomb gap $ \varDelta_{CG} $ = 80 meV. Now to compare these results with the bulk value of manganites in the low concentration region, we plot our resistivity data with the available resistivity data\cite{ref18,ref19} of low doped single crystals in FIG. 3.

 \begin{figure}[t]
   \begin{center}
   \includegraphics[width=8.5cm,height=7cm]{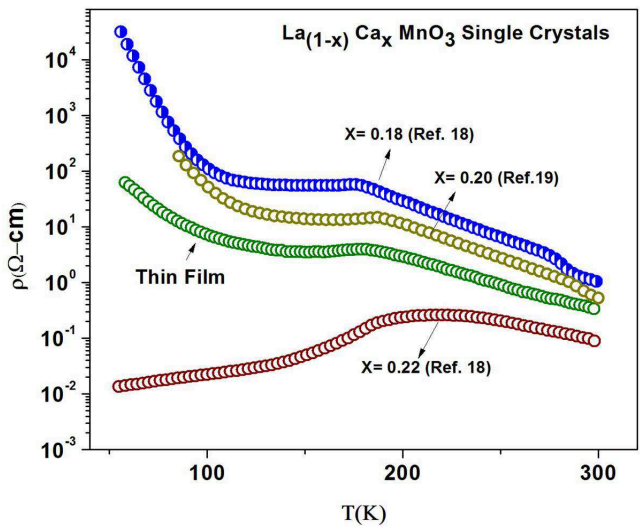}
   \end{center}
   \caption{(Color online)Comparison of resistivity data of the thin film with  different bulk single crystals with various doping ($ X = 0.18 - 0.22 $)}
   \label{Fig3}
   \end{figure}      

From the FIG. 3, we find that the resistivity of the thin film behaves like a single crystal and its magnitude is in between 0.2 and 0.22 doping region that means the composition of the film is changed due to the strain between the film and the substrate. We quantitatively compare the data obtained from $\rho-T$ plot of the film with the available data of single crystals tabulated in TABLE. 1.
   \begin{table}
      \caption{\label{tab:table 1}Comparison of electrical parameters of the thin film with different doping of LCMO}
      \begin{ruledtabular}
      \begin{tabular}{ccccc}
      Parameter& X=0.18& X=0.20& Thin Film\\
      \hline
      $ T_{C}(K) $& 165 & 185 & 181 \\
      $ T_{FMI}(K) $& 107 & 120 & 137\\
      $ \rho_{300}(\Omega.cm) $& 32& 0.78& 0.33 \\
       $ \rho_{56}(\Omega.cm) $&  $ 3\times 10^{4} $& $ 2\times 10^{4}$& 69  \\
       $ \rho_{0}( \Omega.cm K^{-1/2}) $& $ 6\times10^{-6} $&  $ 8\times10^{-6} $&  $ 3\times10^{-4} $ \\
       $ T_{0} (K) $& $ 3\times10^{4} $& $ 3\times10^{4} $& $ 6\times10^{3} $ \\
       $ \varDelta_{CG}(meV)$& 200& 153& 80\\
       $ \xi(nm)$& 0.2& 0.2& 0.5\\
       \end{tabular}
       \end{ruledtabular}
       \end{table}
It is clear from the TABLE.1 that $ T_{C} $ of the film shifts about 16 K to higher temperature from the X=0.18 doped single crystal whereas it 4 K below from the X = 0.20 single crystal. There is also a gradual shift of $T_{FMI}$ to higher temperature for X=0.18 single crystal to the thin film and the change is nearly 30 K. If we compare the resistivity of the three samples than as usual X=0.18 and 0.20 doped single crystals has higher resistivity values at 56 K and 300 K than the film and the $ \rho_{56} $ and $ \rho_{300} $ value of the film is such that it indicates the critical composition region of the film. So strain between film and the substrate not only triggers the FMI state but also changes the room temperature resistivity of the film.  Another important factor is the Coulomb gap  in FMI state and if we compare it for the three samples, we get continuous decrease of this gap up to the film . The localisation length ($\xi$) of the carriers in FMI state determines the conductivity in this sate and it is inversely proportional to the $ T_{0} $.\cite{ref17} As the $ T_{0} $ value of the the thin film is one order less than the X=0.18 and X=0.20 doped single crystals, the localisation length of the film is greater than the other two bulk single crystals and for that the film should be less resistive and indeed it reflects in  the resistivity data of the film (in FIG.3) with comparison to the single crystal values. So  the film shows single crystal like electrical transport behavior and its composition nearly matches with the X=0.21 doped region as the resistivity falls in between the X=0.20 and X=0.22 doping region with Coulomb gap reduced nearly half from the $ \varDelta_{CG} $ value of  the X=0.20 single crystal. 

\subsection {Magnetoresistance of the thin film}
It is well known that MR vanishes in the FMI state of bulk single crystal manganites but there is no such study in the thin film due to the difficulty of retaining the FMI state in the thin film. As it is established that the composition of the film is in critical region and it shows the FMI state, we have done detail MR measurement shown in FIG.4 with varying magnetic field (0-10 T) and temperature (80 K- 250 K). 
  \begin{figure}[h]
        \begin{center}
        \includegraphics[width=9cm,height=7cm]{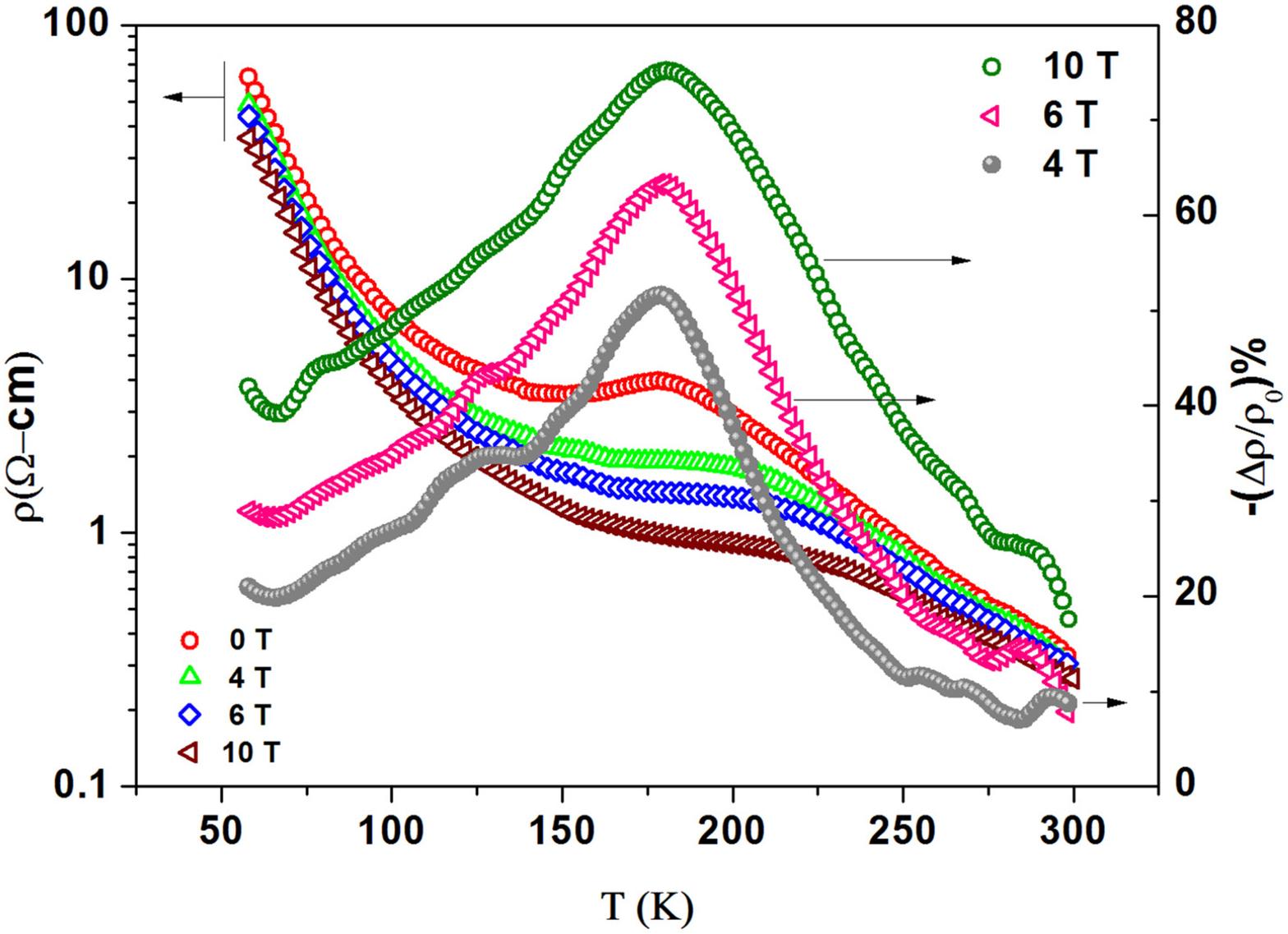}
        \end{center}
        \caption{(Color online) Resistivity measurement  of the sample with different magnetic field. In the right hand side  $ -MR\% = -(\varDelta\rho/\rho_{0})\% $ with temperature is plotted .} 
        \label{Fig4}
        \end{figure}      
 From the FIG. 4 , it is clear that there is a noticeable decrease in $ \rho $ of the film with the application of the external field and the film has negative magnetoresistance($-MR\% $)  in the whole temperature range. We have plotted the variation of $ -MR\% [= -(\varDelta\rho/\rho_{0})\%] $ with T.
 Magnetoresistance of the sample increases with the lowering of temperature but after a certain temperature($T_{MR}$) it collapses in the FMI state of the sample. At the highest applied field (10 Tesla), MR increases to $ 75\% $ and then decreases to $ 40\% $ which is very high with respect to the FMI state of low doped manganites. The magnetoresistance in the FMI state decreases with the lowering of the field and we have observed that at the lowest applied field of 1 T it is nearly $ 5\% $. $ T_{C} $ and $ T_{FMI} $ of the film shifts to higher temperature with the application of the field and the $ T_{MR} $ is in between the $ T_{C} $ and $ T_{FMI} $.  As the low temperature resistivity of the film follow the ES-VRH model magnetoresistance of the film depends upon magnetic length which is expressed as $ \lambda_{m}$ = $(\frac{ch}{2 \pi eH})^{1/2}$  where symbols have their usual meaning with H as an applied field\cite{ref17}. Now if the $ \lambda_{m} >= l_{h}(T,H=0)$, where $ l_{h} $ is the localisation length of the carriers,then no significant magnetoresistance effect will be observed and normally it is found that $ \lambda_{m} $ is greater then $ l_{h} $ in the FMI state \cite{ref18}. MR is quite high in our case than  usual  FMI state and we have calculated $\lambda_{m} $ and $ l_{h} $ at 100 K for T $ < T_{FMI} $. In our case $\lambda_{m} $ = 5 nm while $ l_{h} $ = 1 nm but in case of X=0.18 doped bulk single crystal the ratio between $ \lambda_{m} $ and $ l_{h} $ is large\cite{ref18}. This may be one of the reason for the high MR in the FMI state of the film than the single crystals. We have tabulated  electrical parameters of the film with the applied filed by fitting different $ \rho-T $  curves for different magnetic field with adiabatic polaronic model and listed in the TABLE. 2. 

 \begin{table}
    \caption{\label{tab:table 2}Comparison of electrical parameters of the thin film with different field }
    \begin{ruledtabular}
    \begin{tabular}{ccccc}
    Parameter& 0 T& 4 T& 6 T& 10 T\\
    \hline
    $ T_{C}(K) $& 181& 203& 213& 225 \\
    $ T_{FMI}(K) $& 137& 140& 145& 150 \\
    $ T_{MR}(K) $& -& 178& 179& 181 \\
     $ E_{a}(meV)$& 144& 128& 119& 110\\
     \end{tabular}
     \end{ruledtabular}
     \end{table}
From the TABLE. 2, it is clear that external field decreases the  activation energy for transport which in turn reduces $ \rho $ and increase $-MR\%$ up to $ T_{C} $.
 
 \subsection {Field dependence of the magnetoresistance}
 In our film , we established the presence of three distinct regions i.e. ferromagnetic insulating, ferromagnetic metallic (FMM) and paramagnetic insulating (PI) in the $ \rho-T $ curve. It is interesting to study the field dependence of the magnetoresistance in these regions. We have measured the field dependence MR from 60 K to 275 K shown in FIG. 5.
  \begin{figure}[h]
        \begin{center}
        \includegraphics[width=9cm,height=7cm]{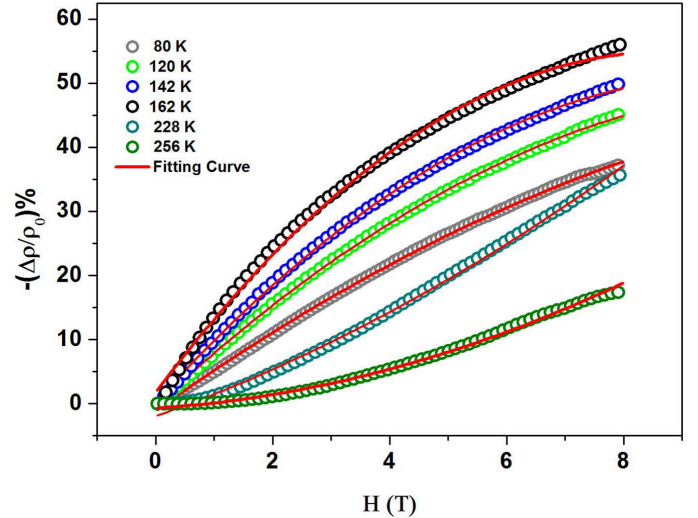}
        \end{center}
        \caption{(Color online) Field dependence of the magnetoresistance with $ -MR\% = aH+bH^{2} $ fit. } 
        \label{Fig 5}
        \end{figure}      
We have chosen the tempeartures for $ T< T_{FMI}$(80, 120 K), $T_{FMI}<T<T_{C}$(142, 162 K) and $T_{C}<T< 300 K$(228, 256 K).Up to 162 K, the nature of dependence of MR on H is same but it change its sign after 202 K. We have fitted the MR vs H curve at different temperature with a equation $ -(\varDelta \rho/\rho_{0})\% = aH+bH^{2} $, where \textquotedblleft a\textquotedblright is the linear term and \textquotedblleft b\textquotedblright is the quadratic term  of the magnetoresistance. The estimated \textquotedblleft a\textquotedblright and \textquotedblleft b\textquotedblright as a function of temperature has been plotted in FIG. 6.
  \begin{figure}[t]
          \begin{center}
          \includegraphics[width=9cm,height=7cm]{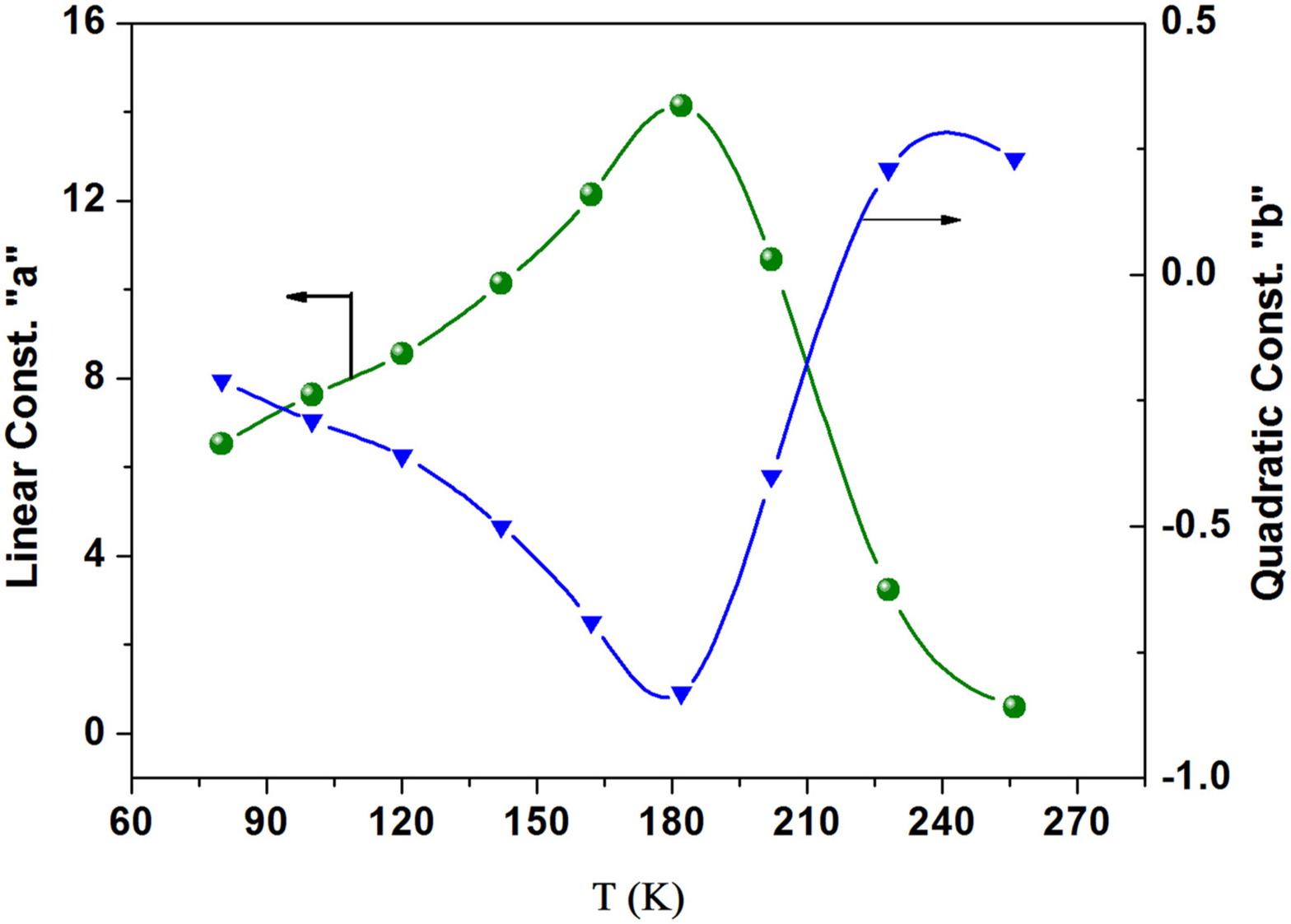}
          \end{center}
          \caption{(Color online)Temperature dependence of the linear and quadratic term } 
          \label{Fig 6}
          \end{figure}       
 We found that linear and quadratic terms  have different magnitude and tempearture dependence. The linear term  is increasing up to 180 K and then it is decreasing with T and always remains positive while the quadratic term changes sign and decreases up to 180 K and after that it increases again. Through out the whole temperature range \textquotedblleft a\textquotedblright is greater than \textquotedblleft b \textquotedblright but \textquotedblleft b\textquotedblright changes its sign at higher tempearture and these two parameters give the parabolic temperature dependence of the MR . The most important feature is that both these parameters show peak at 180 K which is the $ T_{C} $ of our film. 
\subsection {Anisotropic magnetoresistance (AMR) of the film}
Anisotropic magnetoresistance effect (AMR) in manganites is different from the conventional metallic systems\cite{ref20,ref21} and most of the study related to this effect has been done on the strained thin films of manganites in the optimally doped region\cite{ref22}. The lattice strain in the film due to the the lattice mismatch between substrate and the film  generally plays an important role in tuning the AMR.Reduction of film thickness which increase the lattice strain can increases AMR effect.Recently AMR effect is investigated in weakly doped LSMO (x=0.16) thin film and on low doped LCMO(x=0.21) bulk single crystal sample\cite{ref12,ref23}. Now as the resistivity of our film falls into the region between X= 0.20 and 0.22 bulk single crystal that means it is  equivalent with the X=0.21 doped single crystal sample. So we have investigated the AMR effect on this film (FIG.7) with  high magnetic field (8 T) and compared it with respect to the measurement of the  X= 0.21 doped bulk single crystal with 14 KOe field. 
  \begin{figure}[h]
            \begin{center}
            \includegraphics[width=9 cm,height=7cm]{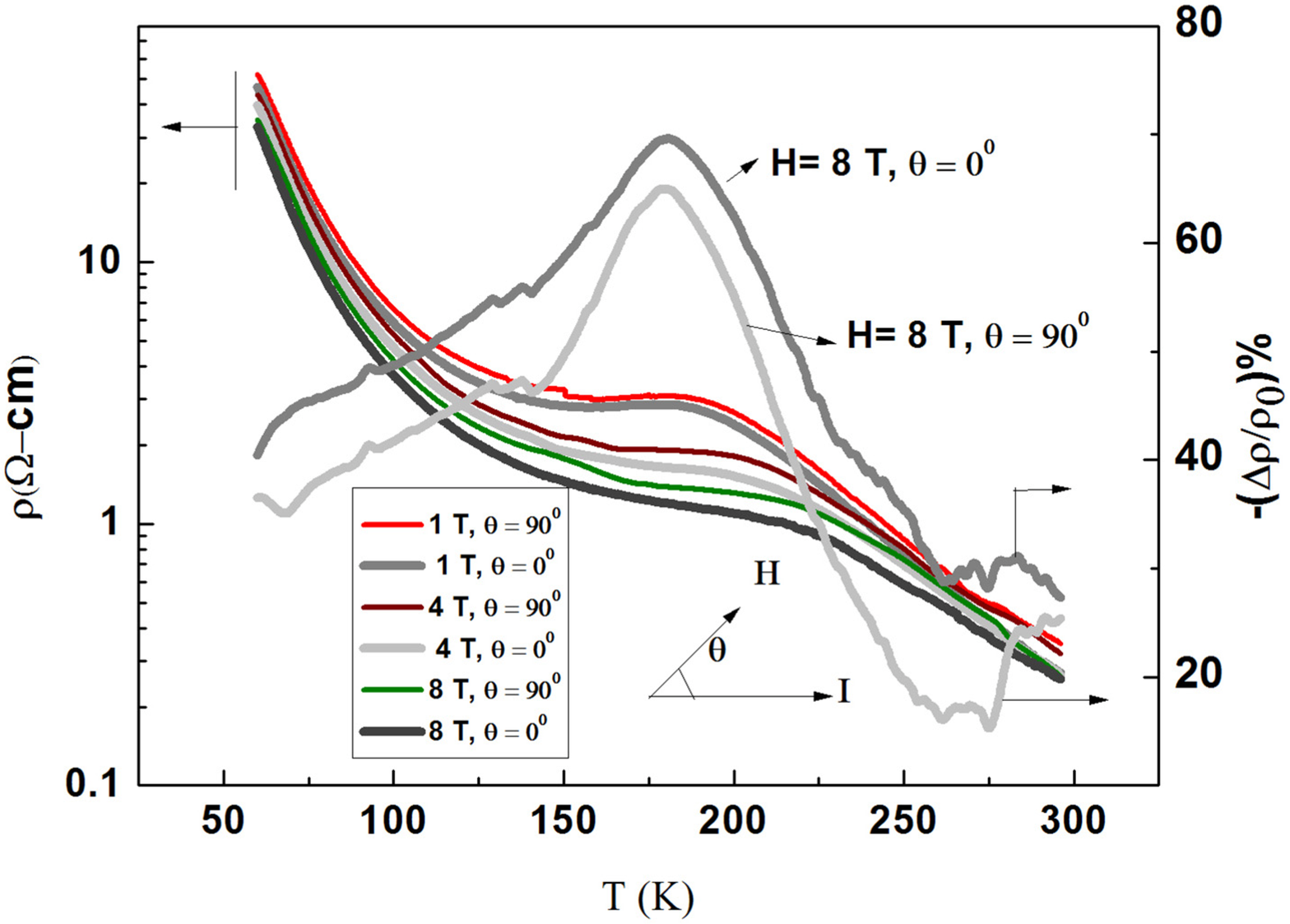}
            \end{center}
            \caption{(Color online) AMR of the film  } 
            \label{Fig 7}
            \end{figure}       
 AMR is defined as the change in resistivity due to the change in current (I) direction with the applied field (H). If $ \theta $ be the angle between magnetic field and current direction then  we have done the resistivity measurement of the film with different field (1-8 T) with $ \theta $ = $ 0^{0} $ and other with $ \theta $ = $90^{0} $ i.e. when field direction is perpendicular to the current direction, however both are in the same plane. Generally in manganites $ \rho _{\theta = 90^{0}}$ is greater than$ \rho _{\theta = 0^{0}}$ while in metal it is reverse. From the FIG. 7, we can see that the $ \rho _{\theta = 90^{0}}$ is greater than $ \rho _{\theta = 0^{0}}$  and  AMR effect is gradually  increasing with the applied field and there is nearly $ 20\% $ decrease in the resistivity of the film at 8T in $ \theta $  = $90^{0} $ direction. We  have calculated the change in magnetoresistance at the highest field due to this effect and plotted in the right side of the graph in FIG. 7. Magnetoresistance is higher for the parallel direction than the perpendicular. Now the other factors which influence the AMR effect is the local uniformity of the material and the increase of the spin polarisation due to the enhancement of the applied field and decrease of tempearture. In our case the change in the MR due to these effect is maximum in the FMI state as well as in the high tempearture phase. So like the X=0.21 bulk single crystal, we have also seen the  prominent AMR effect in this film.  
\section {Conclusion}
In conclusion we have made a single crystal like FMI thin film of manganites in the critically doped region by pulse laser deposition from a X=0.15 doped LCMO pellet. XRD of the film reveals that the film is strained and the electrical transport measurement confirm its single crystal like behavior  and the doping region which is nearly X=0.21. We have done the magneto transport of the film and found that it has a negative magnetoresistance as the bulk single crystal sample and the MR peaks at a certain tempearture ($ T_{MR} $)  and collapses in FMI state but its value is quite high in respect of the other single crystal samples. The $ T_{MR} $ of the sample increases with the increasing field and it is  close to the value of $ T_{C} $ at 0 T field. The $ T_{C} $ and the $ T_{FMI} $ of the sample also increases with the applied fileld but the effect is more prominent in case of $ T_{C} $ which is the metal insulator transition. So like the CMR materials, insulator-metal transition temperature changes with the applied field. Magnetoresistance of the film shows two kinds of field dependence; one is linear and the other is quadratic and they varies differently with the temperature with opposite sign but more interestingly they have a peak at the $ T_{C} $.At the end  there is a noticeable AMR effect in the film  due to the lattice strain and the local inhomogeneity in the film. So we have done a detail investigation of the electrical as well as magneto transport of the single crystal like film in the critically doped region and tried to give some physical insight which has not done before.   

\section{ACKNOWLEDGMENTS}
The authors acknowledge the financial support from the Department of Science and Technology, Government of India and  CSIR. One of the author Rajib Nath would like to thank Dr. Kaustuv das for his useful suggestions and discussions.

\end{document}